# Cyclic thermo-mechanical performance of granular beds: Effect of elastoplasticity

Si Suo[1,2]*, Marigrazia Moscardini[3], Verena Becker[1], Yixiang Gan[2,4]*, and Marc Kamlah[1]

[1] Institute for Applied Materials, Karlsruhe Institute of Technology (KIT), Germany

[2] School of Civil Engineering, The University of Sydney, NSW 2006, Australia

[3] Department of Civil and Industrial Engineering, University of Pisa, 56122 Pisa, Italy

[4] Sydney Nano, The University of Sydney, NSW 2006, Australia

* Corresponding author: si.suo@sydney.edu.au,  yixiang.gan@sydney.edu.au

**Abstract:** Understanding the coupled thermo-mechanical behaviour of compacted granular beds can benefit various industrial applications, such as pebble bed design in fusion reactors. In this study, a thermo-mechanical discrete element method based on our previous work is improved and adapted to investigate the cyclic thermo-mechanical performance of gas-filled granular materials composed of elastoplastic grains. An interparticle contact model is developed considering the plastic deformation of grains. Through the simulation on a representative volume element of beryllium pebble beds, we provide grain-scale insight into the evolution of thermal conductivity and stress. The simulation results suggest that the network of thermal contacts is impeded by plastic deformation leading to a significant drop of thermal conductivity during cooling. This effect can be suppressed by increasing the initial packing factor. Not limited to pebble bed design, the conclusion of this work can also pave the way for optimizing powder-based manufacturing and energy storage, where combined thermo-mechanical loading conditions and elastoplastic deformation of individual particles are involved.

# Highlights

- A modified contact model is developed within a DEM framework, which can precisely describe the mechanical behaviours of compacted granular beds composed of elastoplastic particles.
- Coupled with an improved thermal network model, we build up a comprehensive numerical framework for estimating the thermo-mechanical performance of granular beds.
- Through the grain-scale simulations, we provide a detailed insight into the cyclic behaviour of beryllium pebble beds and highlight the effect of irreversible deformation, e.g., ETC reduction and network resilience in the form of effective stress drop.

# 1. Introduction

Gas filled granular materials are extensively encountered in nature and industrial applications, such as geothermal engineering [1], thermal energy storage [2], and chemical reactors [3]. Especially for the breeder blankets in fusion reactors [4, 5], the helium cooled pebble bed (HCPB) is a potential option. For a closely packed pebble bed, the heat transfer process is generally coupled with mechanical behaviours, especially during heating-cooling cycles. Due to lack of grain-scale insights, the widely-adopted effective medium theory [6] is not applicable to such complex dynamical processes. Therefore, the design and management of compacted granular beds necessitate a further understanding of effects of thermal-mechanical interactions on their performance.

Since a gas-solid granular assembly is featured by multiple phases and heterogeneity, the heat transfer process is dominated by manifold factors including bulk material properties such as thermal conductivity and heat capacity of each phase, microstructure features such as particle size and spatial topology, e.g., relative position of particles and local clustering, and environmental conditions like temperature and gas pressure. Thus, there have been plenty of experimental works [7-9], numerical studies [10-14] and theoretical analyses [15-17] on obtaining the effective thermal conductivity (ETC). Among them, the DEM-based numerical framework, i.e., the thermal DEM, is proved to be a promising tool and has been widely adopted since its core concept captures the discrete nature of granular materials [18-21]. In thermal DEM, a granular assembly is characterized by a resistor network, i.e., nodes correspond to particles between which the link represents the thermal resistance. Based on different assumptions on the heat flux pathways, some grain-scale analytical solutions have been developed, such as a gas film model [22], Voronoi cell based models [10, 23, 24] and the Batchelor and O'Brien analytical solution [25], to evaluate the inter-particle thermal resistance. Recently, combined with other techniques, e.g., machine learning, the thermal DEM has shown the potential to be applied to multi-scale investigations [26, 27]. Most of the existing models for describing interparticle thermal conductance, however, focused on the elastic mechanical contacts for the sake of simplicity, leading to underestimate ETC and ignore the irreversible behaviour due to possible plastic particle deformation during cyclic loading.

Although existing research has obtained essential achievements on the static characterization of heat transfer in a compacted granular bed, further study on the state evolution is necessary for a more exact and comprehensive evaluation of its thermal performance since such process is dynamical and strongly coupled with the mechanical state, especially for the granular materials working in a complex and extreme environment. For instance, the pebbles in a confined breeder unit are cyclically heated and cooled with neutron irradiation turned on and off [28]. The resultant thermo-mechanical coupling of compacted granular materials may lead to appealing physics. Chen et al. [29] discovered that an unconsolidated granular assembly can also be packed by thermal cycling besides traditional ways like vibration or tapping. With the help of thermo-mechanical DEM simulations, Vargas et al. [30] provided a grain-scale insight on how the evolution of force chains is influenced by the thermal expansion and pointed out that the thermal packing phenomena shown in [29] can be described by the Mittag-Leffler law. Furthermore, Divoux et al. [31] investigated the long-term thermal packing in a granular pile and demonstrated that the compaction mode can be controlled by the temperature amplitude. Other related topics, such as thermal effects on stress and rigidity of assemblies, can be found in [32-34]. Nevertheless, the current discussion is limited within the elasticity regime i.e., residual deformation at grain scale is ignored though the effects of plasticity are commonly reported in certain works [35-37]. Besides, most of the above works focus on the thermally induced mechanical response only, while knowledge of how the change of microstructure resulting from the thermal cycling inversely influences the thermal conductivity is scarce. In a granular bed under cyclic loading, the irreversible deformation can be a combination of irreversible changes of the topological packing structure [38] and plastic deformation within the grains. Especially for plastic grains like metal pebbles, the long-term performance of such granular materials should be carefully investigated due to the irreversible deformation that accumulates with thermal cycling.

In this work, a DEM approach implemented in the KIT in-house code is adopted to investigate the thermo-mechanical behaviours of compacted granular beds during heating-cooling cycles, originating from a representative scenario of a fusion breeder blanket unit. The numerical framework based on our previous works [32, 39, 40] is further improved. Specifically, an elastoplastic contact model suitable for a large range of plastic deformation is developed and validated in Section 2.1, so that the current framework can be applied to metal pebble beds. The thermal conductance between neighbouring particles is thoroughly explained in Section 2.2. The Smoluchowski effect is expressed in a more physical way since separate formula for thermal contributions of each phase are adopted here instead of a combined one in [39], to better capture the thermal conductance at low gas pressure region. Representative volume element (RVE) models are extracted from a beryllium pebble bed, as a typical granular material composed of elastoplastic grains [41], and described in Section 2.3. In Section 3, the influence of plasticity and packing factor on the cyclic performance of pebble beds, i.e., the evolution of ETC and thermal stress, are discussed. Finally, some conclusions are made in Section 4.

## 2. Numerical modelling

### 2.1 Modified contact model considering plastic deformation

The mechanical normal contact behaviour during the compression of two particles may cross different material phases. In the original Thornton theory [42], the grain material keeps elastic with normal contact behaviour defined by Hertzian theory at the initial stage. With increasing compression, the grain material begins yielding when the contact pressure reaches a certain limiting yield pressure $p_y$, which is suggested to be constantly proportional to yield stress $\sigma_y$ in [42], i.e., $p_y = \tau\sigma_y$ ($\tau = 1.6\sim3.0$). Furthermore, the pressure distribution is based on the Hertzian solution, however it is truncated such that the central part is replaced by a constant value of $p_y$. If unloading occurs after yield, the normal contact behaviour is calculated by the Hertzian solution considering a shift such that the normal contact force vanishes at an irreversible overlap $\delta_{np}$. Similarly, the normal contact force increases along the unloading path when reloading occurs, and the plastic compression continues beyond the unloading point ($\delta_{n,max}$, $F_{n,max}$). In summary, the normal contact force between two touching particles, $i$ and $j$, during elastic loading, plastic loading, and elastic-plastic (re)unloading can be mathematically expressed as

$$F_n = \begin{cases} \frac{4}{3}E^*\sqrt{R^*\delta_n^3} & \delta_n \leq \delta_{ny} \\ F_{ny} + \pi R^* p_y(\delta_n - \delta_{ny}) & \delta_n > \delta_{ny}, \dot{\delta}_n > 0 \\ \frac{4}{3}E^*\sqrt{R_p^*(\delta_n - \delta_{np})^3} & \delta_n > \delta_{ny}, \dot{\delta}_n < 0 \end{cases}, \quad (1)$$

$$\text{where } F_{ny} = \frac{4}{3}E^*\sqrt{R^*\delta_{ny}^3}, \quad (2)$$

$$\delta_{ny} = \left(\frac{\pi p_y}{2E^*}\right)^2 R^* \quad (3)$$

$$R_p^* = \frac{4}{3}\frac{E^*}{F_{n,max}}\sqrt{\left(\frac{2F_{n,max} + F_{ny}}{2\pi p_y}\right)^3}, \text{ and} \quad (4)$$

$$\delta_{np} = \delta_{n,max} - \left(\frac{3F_{n,max}}{4E^*\sqrt{R_p^*}}\right)^{\frac{2}{3}}. \quad (5)$$

Here, $E^* = \frac{E_i E_j}{E_i(1-\nu_j) + E_j(1-\nu_i)}$ is the Effective modulus, and $R^* = \frac{R_i R_j}{R_i + R_j}$ the effective radius, with $\nu$ as the Poisson ratio. To verify the applicability of this model to beryllium particles, a FEM analysis is conducted on the particle-rigid wall setting as shown in Figure 1(a), where the material behaviour is assumed to be simply linear elasticity followed by perfect plasticity. Exploiting symmetry, this setting represents the compression of two equal spheres. The material parameters are given in the caption of Figure 1 [53]. Compared with FEM results, as shown in Figure 1(b), the Thornton model shows good agreement within a small range of plastic strains. However, with the overlap $\delta_n$ increasing, the assumption of "constant $\tau$" leads to an underestimation of the normal contact force, and moreover adjusting the constant parameter $\tau$ cannot improve the overall exactness. The curve of contact force vs. overlap in Figure 1(b) shows a progressive increasing during the initial plastic loading and then tends to be linear, correspondingly suggesting that the value of $\tau$ varied abruptly with the overlap increasing at the beginning and keeps a constant afterwards. Therefore, we propose here an improved model to predict the normal contact force exactly within a large range of overlap. Instead of a constant $\tau$, we assume $\tau$ varies with the evolution of overlap and introduce a parameter $\tau_{mat}$ according to,

$$\tau_{mat} = \frac{2E^*}{\pi\sigma_y}\sqrt{\frac{\delta_{n,max}}{R^*}}, \text{ and} \tag{6}$$

$$\tau = 1 + k\left(\frac{\tau_{mat}-1}{\tau_{mat}+1}\right)^n, \tag{7}$$

where $k$ and $n$ are two fitting parameters. In Eq. (7), $\tau_{mat} = 1$ and thus $\tau = 1$, when $\delta_{n,max} = \delta_{ny}$, where the material begins plastic deformation. Furthermore, $\tau_{mat} \to \infty$ for $\delta_{n,max} \to \infty$, and thus, $\tau$ tends to the maximum and platform value of $k + 1$ for increasing overlap. Therefore, how fast the transition from progressive to linear behaviour occurs can be characterized by the index $n$. After nonlinear regression analysis based on FEM analysis results, the fitting parameters $k = 6.66$ and $n = 5.0$ in the proposed model are obtained with $R^2 = 0.95$. As shown in Figure 1(c), the modification in Eq. (7) enables the Thornton model effective not only for a wide range of compression overlap. Furthermore, as the comparison to FEM shows, the modified model also covers various working temperatures through the temperature dependence of the material parameters as it is given by the material database [53] for elasticity and plasticity, e.g., $\sigma_y$ and $E$ in dependence on temperature. The proposed model has the potential for predicting the mechanical behaviour of various plastic materials while the current fitting parameter set is only suitable for soft metals, i.e., the yield strain is smaller than 0.5%, or $\sigma_y/E \leq 0.5\%$.

For the tangential contact behaviour, a simplified Mindlin-Deresiewicz model is used here in combination with the Coulomb friction scheme [43], i.e., the tangential force $\boldsymbol{F}_t$ can be updated as

$$\Delta \boldsymbol{F}_t = 8G^* \sqrt{R^* d} \Delta \boldsymbol{\delta}_t, \text{ and} \tag{8}$$

$$F_t = \min(\mu F_n, |\boldsymbol{F}_t + \Delta \boldsymbol{F}_t|), \tag{9}$$

where $G^* = \left(\frac{2-\nu^i}{G^i} - \frac{2-\nu^j}{G^j}\right)^{-1}$ is the effective shear modulus with the shear modulus of each material $G^k = \frac{E^k}{2(1+\nu^k)}$, $k = i, j$, and $\mu$ is the friction coefficient. So far, in this numerical scheme $\mu$ is the only one undetermined parameter since it is related to particle surfaces while other parameters including Young's modulus $E$, Poisson's ratio $\nu$, and yield stress $\sigma_y$ can be obtained by the material test. To further investigate the performance of the proposed plastic contact model at meso scale, a uniaxial compression test on a cubic RVE which assembles 5000 mono-sized particles randomly distributed in a periodic region [44] with an initial packing factor of 63% is simulated and validated against experimental results [36]. As shown in Figure 1(d), with the friction coefficient $\mu$ increasing, the assembly tends to be stiffer since the friction-induced tangential contact force between particles also contributes to the average coordination number as well as to the global stress significantly [45, 46]. The case with $\mu = 0.03$ well reproduces the stress-strain curve at loading stage observed in the experiment, while at the unloading stage the curve shape can be captured by the present model though there is a certain parallel shift compared with experimental data. This may result from material creeping at the transition from loading to unloading which may be suspected from the smooth curve shape in the experiment during the transition.

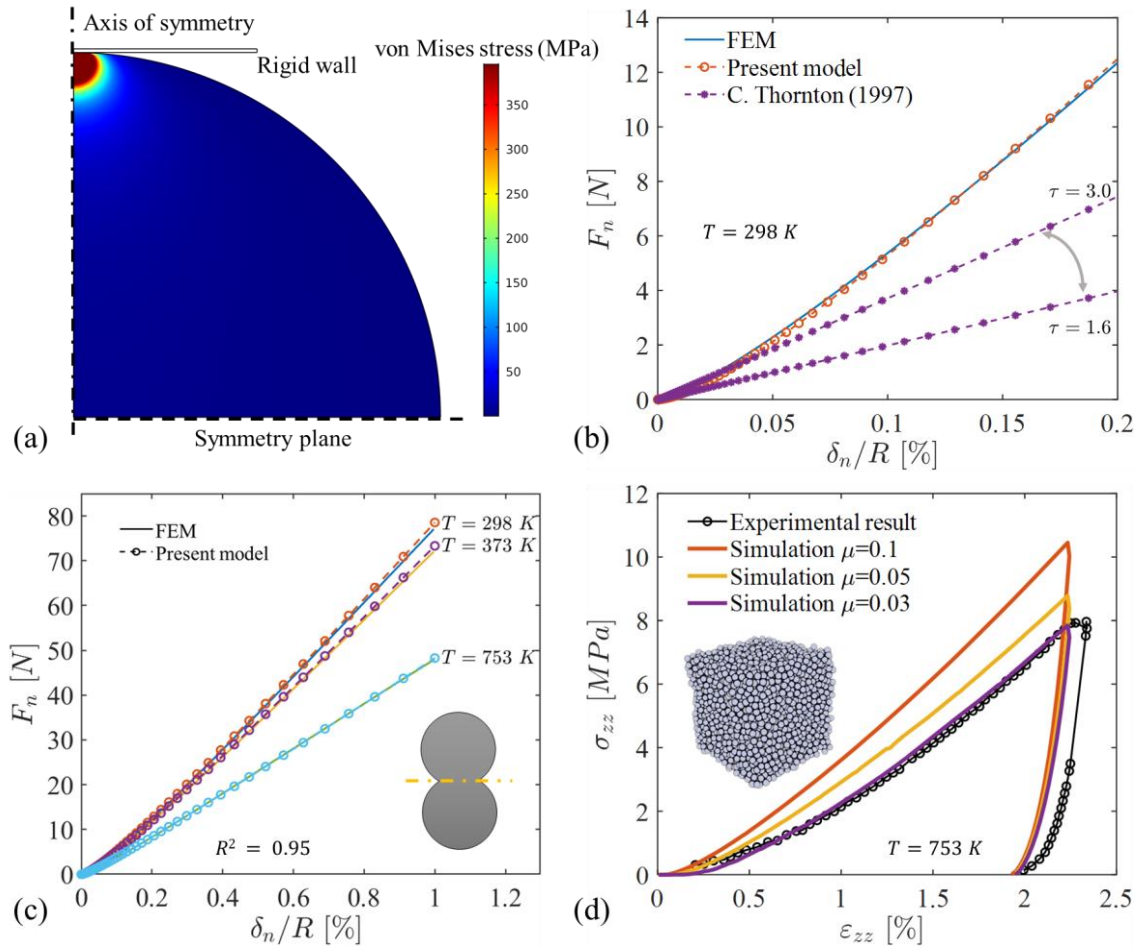

Figure 1. (a) FEM model; (b) comparison among FEM result, original Thornton model and present model; (c) validation of the present model under different temperatures against FEM analysis with (T = 298 K, $\sigma_y$ = 395.51 MPa and E = 276.65 GPa; T = 373 K, $\sigma_y$ = 364.80 MPa and E = 272.70 GPa; T = 753 K, $\sigma_y$ = 226.48 MPa and E = 252.71 GPa [53]); (d) validation of DEM simualtion against experimental data [47].

## 2.2 Thermal network model

In gas filled granular materials, the mechanisms of heat transfer consist of heat conduction, convection and radiation. However, for a closely packed granular material under limited gas pressure (<0.15 MPa [41]), the gas flow is so slow that the corresponding convection can be ignored. Meanwhile, some works [10, 48] on heat radiation in granular materials suggest its contribution to the effective thermal conductivity can be neglected especially for dense packing. Therefore, we develop the thermal network based on the Batchelor and O'Brien analytical solution [25] considering the particle-particle or particle-gas-particle conduction and the Smoluchowski effect [49]. The effectiveness of this conceptual model has been validated in the previous work [39], and here we modify the model reported in [12] to calrify the phase contribution (fluid and solid) separately. Specifically, as shown in Figure 2(a), the thermal conductance $C_{ij}$ between two particles ($i$ and $j$) can be simplified as three thermal resistances in parallel, i.e.,

$$\frac{1}{C_{ij}} = \frac{1}{C_i} + \frac{1}{C_{ct}} + \frac{1}{C_j}, \text{ where} \tag{10}$$

$$C_n = \frac{4\pi K_s (\chi R^*)^2}{R_n}, (n = i \text{ or } j), \text{ and} \tag{11}$$

$$C_{ct} = C_{ct}^s + C_{ct}^g, \tag{12}$$

Here $K_s$ is the thermal conductivity of solid phase. In Eq. (12), $C_{ct}^s$ and $C_{ct}^g$ are thermal conductances of solid and gas, respectively. For two touching particles with the overlap $\delta_n$ and contact radius $r_c$, the thermal conductances are given by

$$C_{ct}^s = 2K_s r_c, \text{ and} \tag{13}$$

$$C_{ct}^g = 2\pi K_g R^* [\Delta H_g + ln(\alpha^2)], \tag{14}$$

where $\Delta H_g = -0.05\beta_{ij}^2$ when $\beta_{ij} < 1$, $\Delta H_g = -ln(\beta_{ij}^2)$ when $\beta_{ij} > 100$, otherwise, $\Delta H_g$ is linearly interpolated, and $\beta_{ij} = \frac{\alpha r_c}{2R^*}$, and $\alpha = \frac{K_s}{K_g}$; The calculation of contact radius $r_c$ accounting for elastoplastic deformation can be found in **Appendix A**. For two neighbouring particles with a gap $\bar{\delta}_n$, the thermal conductances follow to

$$C_{ct}^s = 0, \text{ and} \tag{15}$$

$$C_{ct}^g = 2\pi K_g R^* \left[(1 - \lambda_{ij}) \ln(\alpha^2) + \lambda_{ij} \ln\left(1 + \chi^2 \frac{R_{ij}}{h_{ij}}\right)\right], \tag{16}$$

where $\lambda_{ij} = \frac{\alpha^2 \bar{\delta}_n}{2R^*}$ and $\lambda_{ij} \leq 1$. The thermal conductivity of the interstitial gas $K_g$ is a function of the local Knudsen number, thus, accounting for the Smoluchowski effect. For the implementation of the Smoluchowski effect, it can be refered to [39]. So far, the thermal network model is built up and then calibrated by setting the only fitting parameter $\chi$ to match our experimental results [7]. Parameter $\chi$ describes the cross-sectional area of the region of gas and solid thermal transport between two particles. As shown in Figure 2(b), the present model can reproduce the experimental observation when $\chi = 0.69$.

Notably, the decoupling of gas and solid contributions in Eqs. (13~16), which is slightly different from the expressions of the previous one [12], guarantees that the Smoluchowski effect can only function on the contribution of the interstitial gas rather than the whole contact conductance. Specifically, it is assumed in the previous model that $C_{ct}^s$ is a function of $\beta_{ij}$ and implicitly related to the gas pressure. When $\beta_{ij} < 1$, $C_{ct}^s$ is underestimated as $0.44\pi K_g R^* \beta_{ij}^2$ [12] compared with the Eq. (13). So, as shown in Figure 2(c), the prediction of the previous model is overall below the experimental observation [50] while the present model can better capture the influence of varying gas pressure on the effective thermal conductivity of pebble beds.

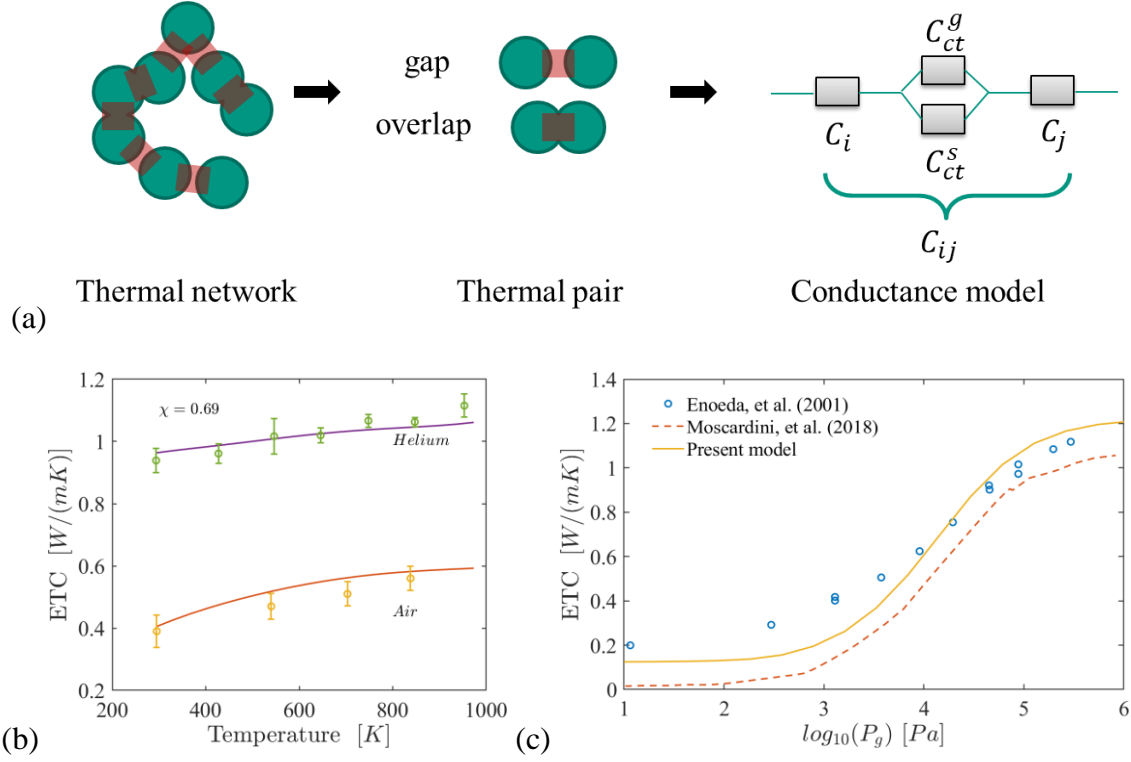

Figure 2. (a) Schematic of thermal network model for pebble beds; (b) Calibration of present model; (c) Comparison among experimental results (previous and present model).

## 2.3 Simulation of cyclic thermal loadings

Due to their granular nature, pebble beds demonstrate strong thermo-mechanical coupling, i.e., stress-dependent thermal conductivity [51], as well as temperature-induced stress evolution [29]. To further clarify the coupling effect on thermo-mechanical performance, especially considering the plastic deformation at grain scale, a 3D RVE, as shown in Figure 3(a), is adopted to mimic a cutout of the beryllium pebble bed in the breeder blanket as a compulsory component in a fusion reactor, where the heat is generated by the neutron irradiation and transferred through pebbles and purge gas [28]. Thus, at the grain scale, the temperature evolution for particle $i$ can be expressed as

$$\dot{T}_i = \frac{1}{m_i c_p}\left(\sum_j C_{ij}(T_j - T_i) + \Psi V_i\right), \tag{17}$$

where $m_i$ and $V_i$ are particle mass and volume; $c_p$ is heat capacity and $\Psi$ is the power density of the heat source. The parameters regarding geometry and working conditions are given in [41]. In addition to the material properties as functions of temperature, as shown in **Appendix B**, the change of particle radius $\Delta R$ due to thermal expansion is also incorporated as

$$\Delta R = R\alpha(T)\Delta T, \qquad (18)$$

where α is the coefficient of linear thermal expansion.

To understand the performance of pebble beds during thermal cycling, the whole simulation consists of the following stages:

1. Packing: Particles of initial radius $R_0$ are randomly packed between two rigid walls at temperature $T_0$.
2. Pre-heating: The temperature of top and bottom wall is fixed as $T_w$ and particles are heated and expanded until the temperature gradient disappears.
3. Heating: Turn on the neutron irradiation, i.e., add the heat source $\Psi$, so that particles are still heated until a steady status is reached.
4. Cooling: Turn off the neutron irradiation, i.e., cancel the heat source $\Psi$, to cool the particles.
5. Finally, the heating and cooling stages repeat another two times.

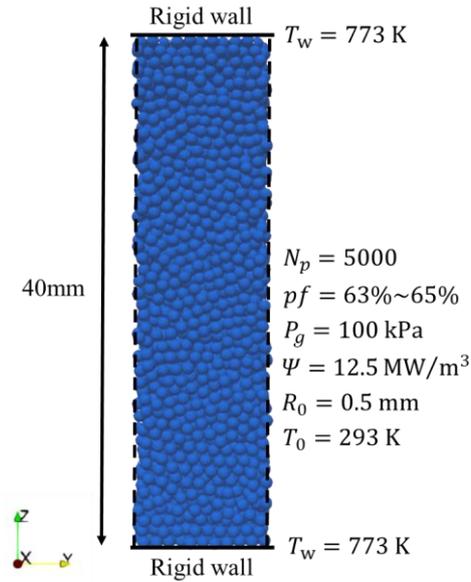

Figure 3. The schematic of calculation model, i.e., 5000 particles are confined between two walls along $z$-axis and periodic arrangement along $x$-and y-axis; for the heat transfer, the rigid walls are set as temperature boundaries and the lateral sides as periodic boundaries.

## 3. Results and discussion

To clarify the influence of plasticity and packing factor on the performance of pebble beds during the thermal cycling, three samples with different packing factors, i.e., 63%, 64% and 65% are tested, for each case considering the particle plasticity or not.

### 3.1 Influence of plasticity

As shown in Figure 4(a) and (b), the ETC and the overall stress $\sigma_{zz}$ in vertical direction follow a similar trend during thermal cycling. During a heating stage, $\sigma_{zz}$ as well as the ETC increases to a maximum when the steady state is reached. Notably, the thermal stress is much smaller when particle plasticity is considered. Although it is hardly visible in Figures 4(a) and (b), it can be inferred from the temperature profile, as shown in Figure 4(c), that the steady ETC of the plastic case is slightly smaller since its peak temperature is a little higher compared with the elastic case. At the end of the heating stage, the heat transfer in a closely packed granular material mainly depends on the solid phase, i.e., most heat flux crosses the particle contact area instead of the interstitial gas, as suggested in Chapter 3.2. According to the contact area distribution shown in Figure 4(d), the particle plasticity results in an obvious shift to smaller contact radius and thus a slightly lower ETC. According to the contact area distribution shown in Figure 4(d), particle plasticity results in an obvious shift to smaller contact radius and thus a slightly lower ETC. This is due to the lower normal stress reached inside the assembly of plastic pebbles compared to elastic ones undergoing the same thermal load.

During the subsequent cooling stage, the overall stress $\sigma_{zz}$ rapidly drops and it reaches almost zero values in the case of particle plasticity. Correspondingly, the contact forces at grain scale disappear due to previous plastic deformation. As will be discussed in the following, plastic unloading during cooling results in less motion of particles back to the position before heating and thus limited resilience of network, and the larger the plastic deformation is, the more particle movement is limited. Regarding the network resilience, it can be locally measured with respect to the particle vertical displacement setting position at the end of the pre-heating stage as a reference where the particle network is expected to be restored to this state after cooling if no plastic deformation occurs. Also, the average of such relative displacement over particles with similar height is an indicator for residual deformation at mesoscale. Compared with the elastic case as shown in Figure 4(e), there exists an evident residual particle displacement for the plastic case as shown in Figure 4(f) at the end of cooling stage, though its particle movement induced by the previous heating is much smaller. Furthermore, such residual deformation is accumulated around up and down sides of the middle layer due to the structural symmetry and the non-uniform heating and cooling along the $z$-axis. Thus, some microscopic properties of the pebble beds also vary from top to bottom, such as coordination number shown in Figure 4(g). Specifically, with the cooling processing, the coordination number around the middle layer is obviously smaller than that in the top and bottom layers and a loss of symmetry is also observed which we assume to be the case due to the plastic contacts being more sensitive to the local non-uniformity; while for the elastic case only global decrease of coordination number can be observed in Figure 4(h) and, overall, the coordination number stays at a significantly higher level.

This plasticity-induced loss of solid particle contacts can lead to another consequence, i.e., the ETC reduction as shown in Figure 4(b). This means that the ETC decreases to a lower level than that at the end of pre-heating stage, because the particles cannot be driven to their previous position since contact forces among particles are very small and even disappear during the plastic unloading. By contrast, both indexes, vertical displacement and coordination number, recover for the elastic case to the original state after each heating-cooling cycle.

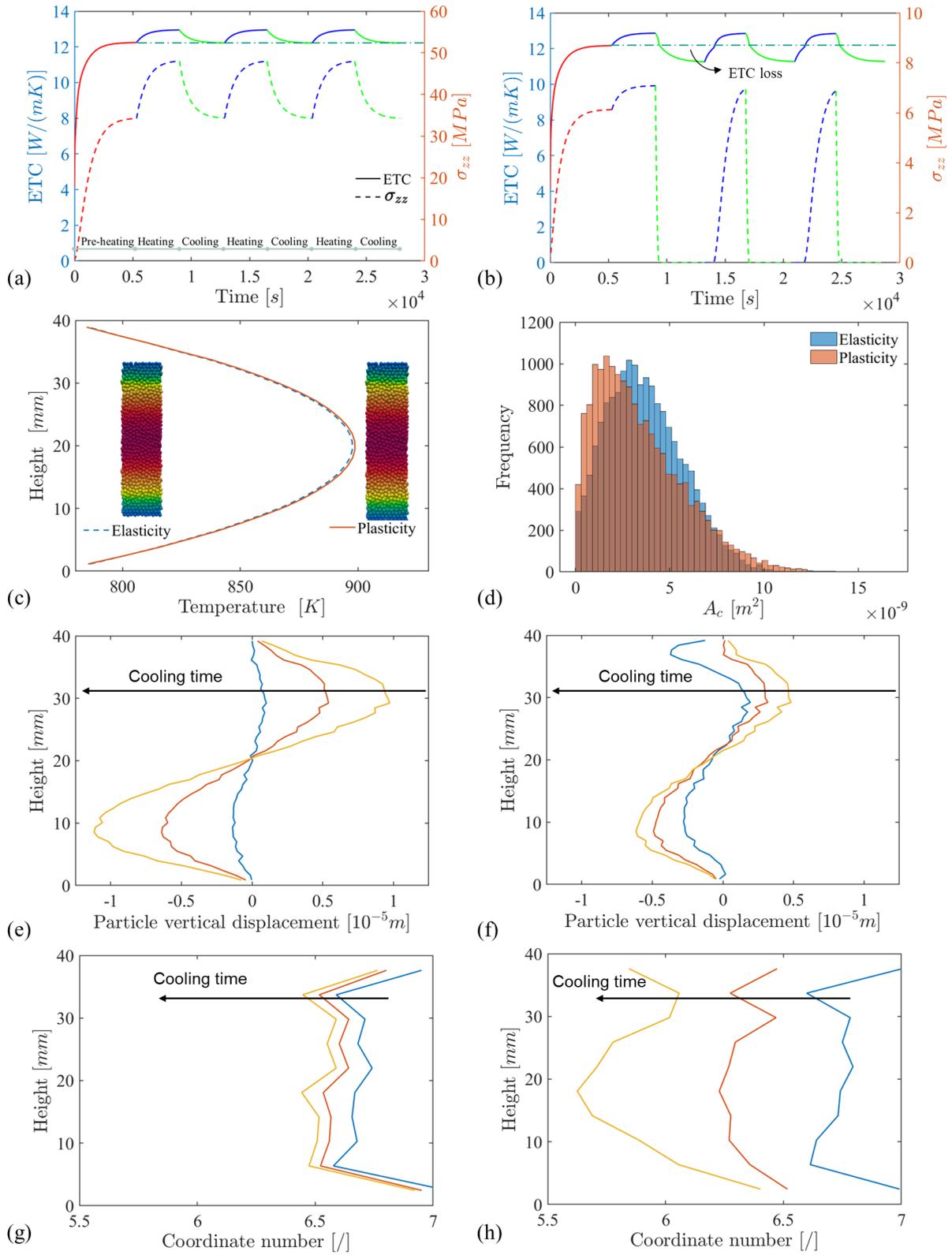

Figure 4. Evolution of ETC and $\sigma_{zz}$ for cases with packing factor of 64% considering perfect elastic (a) or elastoplastic material behavior(b); temperature profile (c) and histogram of contact area (d) for both cases at the end of heating stage; particle vertical displacement and coordination number along $z$-axis for elastic case (e)(g) and plastic case (f)(h) during cooling stage.

## 3.2 Influence of initial packing factor

The packing factor $\phi$ characterizes the amount of solid phase involved in the pebble beds and directly determines their response during heating-cooling cycles. To better understand the influence of initial packing factor, simulations on elastoplastic pebble beds with different $\phi$, i.e., 0.63, 0.64 and 0.65 under the same thermal loops are conducted. As shown in Figure 5(a) and (b), the overall thermal conductivity is enhanced with the packing factor increasing, and correspondingly the thermal stress is also enlarged during the heating stage, since the closer packing structure leads to a larger number of contacts among particles and thus larger contact area so that the heat flux that transfers through the solid phase accounts more. To clarify this point, we adopt two separate thermal networks, as shown in Figure 6(a), i.e., a solid network where the neighboring nodes are linked only by $C_{ct}^s$ in Eqs. (13) and (15), and a gas network where neighboring nodes are linked only by $C_{ct}^g$ in Eqs. (14) and (16). We estimate the phase contribution $I_s$ and $I_g$ during thermal cycling by ratios defined as

$$I_s = \frac{Q_s}{Q_s + Q_g}, \text{ and} \tag{19}$$

$$I_g = \frac{Q_g}{Q_s + Q_g}, \tag{20}$$

Here $Q_s$ and $Q_g$ are the total boundary flux at the rigid walls of the solid and gas network separately under the same temperature difference $\Delta T$. As shown in Figure 6(b), the solid phase always dominates the heat transfer in pebble beds since its contribution ratio is larger than 0.5, though its value is wavy during the thermal cycling. The wavy magnitude is controlled by the packing factor, and specifically $I_s$ for $\phi = 0.65$ a constant level around 0.57 is kept even during the cooling stage while for $\phi = 0.63$ it drops from 0.53 to around 0.50 at the end of cooling stage.

Notably, the ETC reduction during cooling is mitigated to certain extent for a case with higher packing factor since the denser packing limits the residual displacement of particles. As compared in Figure 5(c), the residual particle displacement is limited below $2.5 \times 10^{-3}$ mm for $\phi = 0.65$ so that after cooling the coordination number as shown in Figure 5(d) stays at a high level (>6), i.e., the whole assembly is always at a jamming state [45].

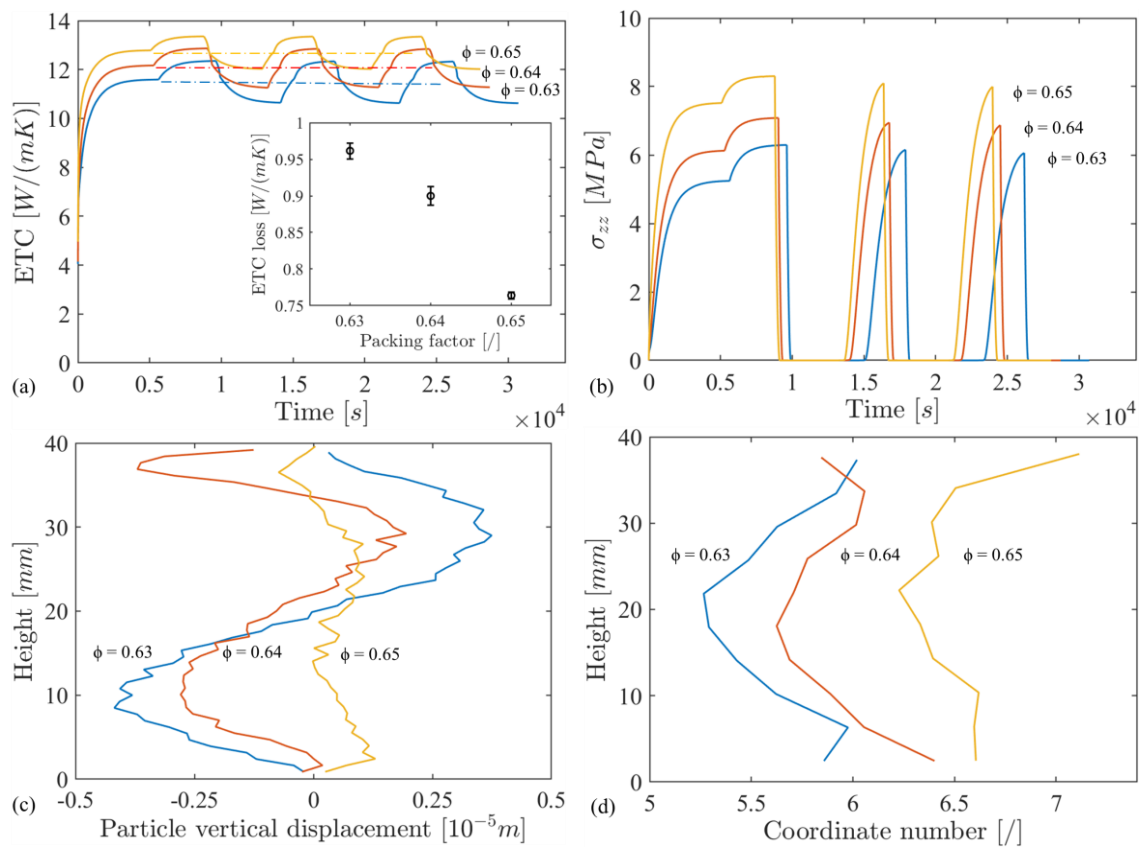

Figure 5. Comparison of the evolution of ETC (a) and $\sigma_{zz}$ (b) and the profile of residual particle vertical displacement (c) and coordination number (d) at the end of cooling stage among elastoplastic pebble beds with different packing factors, i.e., 0.63, 0.64, 0.65. The subfigure in (a) shows the amount of ETC reduction upon cooling with the packing factors.

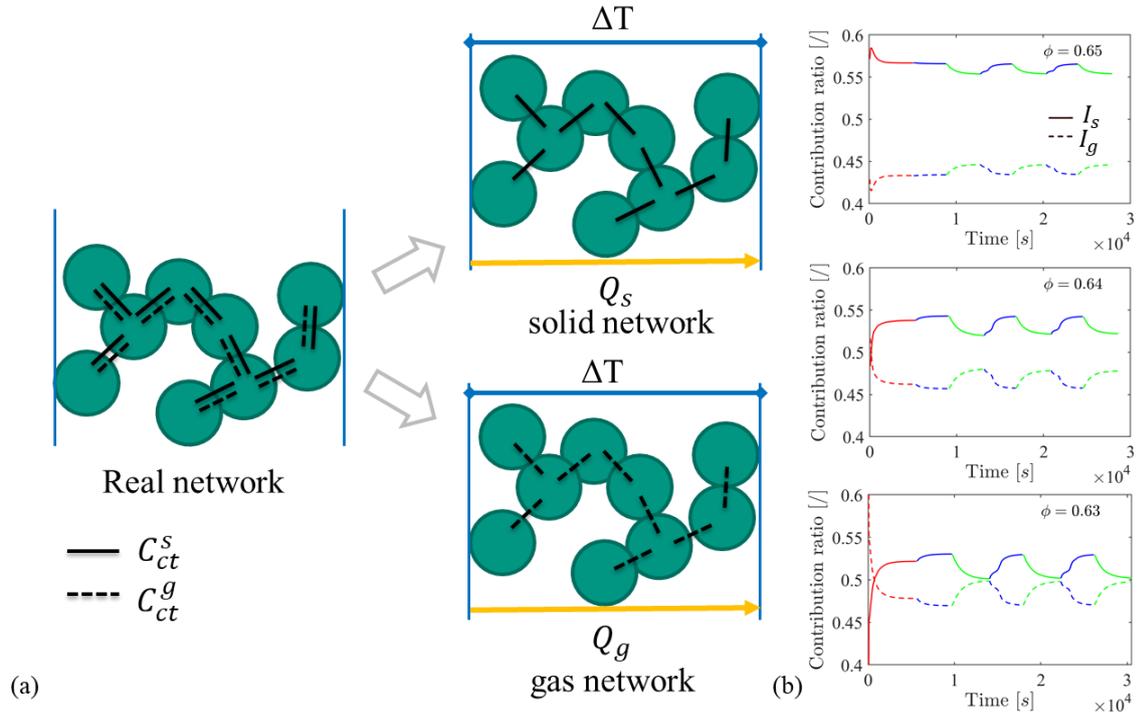

Figure 6. (a) A schematic of separate thermal networks for the estimation of phase contribution; (b) the evolution of phase contribution for cases with various packing factors, i.e., 0.63, 0.64, 0.65.

In summary, the plastic deformation at grain scale can limit network resilience and, therefore, result in the reduction of the ETC during a cooling stage. Since the thermal conductivity of pebble beds is mainly determined by the solid phase, increasing the packing factor can not only enhance the ETC but also mitigate the ETC reduction.

## 4. Conclusion

An exact performance estimation is essential for designing pebble beds operating under complex and extreme loading conditions. In this work, we developed a numerical framework considering comprehensive ranges of thermo-mechanical conditions including various materials, gas pressures and environmental temperatures. In particular:

1) An improved contact model is proposed considering the particle plasticity, and it is validated against FEM analysis at grain scale and against experimental results at RVE scale;
2) A thermal network model based on our previous work is modified by separating phase contributions to the thermal conductance, and the advantage of such modification is proved especially at the regime of low gas pressure.

Through simulations on beryllium pebble beds in the HCPB blanket, we investigated the behaviour of pebble beds during heating-cooling cycles at the grain scale. The simulation results suggest that the thermo-mechanical behaviour of pebble beds depends on the initial packing structures, especially when particle plasticity is considered. Specifically, particle plasticity can lead to the ETC reduction during cooling since the resilience of assembly network is limited by plastic deformation, and such ETC reduction can be suppressed through increasing the initial packing factor.

This work not only sheds light on the effects of thermo-mechanical interactions on the performance of pebble beds, but also presents potential on their optimized design. Furthermore, this work's analysis framework and conclusions can also benefit general powder-based applications involving thermo-mechanical couplings, such as additive manufacturing and thermal management.

**Appendix A: Estimation of contact radius**

Based on the prototype proposed in [52], we rebuild a calculation model of contact radius which is adapted to the proposed plasticity model in Chapter 2.1 as follows:

in the elastic regime,

$$r_c = \sqrt{R^* \delta_n}; \tag{A1}$$

during plastic loading,

$$r_c = \sqrt{R^*\delta_n} \cdot \sqrt{1 + f(\delta_n)}, \tag{A2}$$

where $f(\delta_n) = c\left(\frac{\tau_{mat}-\tau_0}{\tau_{mat}-\tau_0}\right)^b$;

during plastic unloading,

$$r_c = \sqrt{R^*\delta_n} \cdot \sqrt{1 + f(\delta_n)} \cdot g(\delta_n), \tag{A3}$$

where $g(\delta_n) = \left[\left(\frac{\tau_n-\tau_p}{\tau_{max}-\tau_p}\right)^{m_1} - \left(\frac{\tau_{max}-\tau_n}{\tau_{max}-\tau_p}\right)^{m_2} + 1\right]/2$, $\tau_n = \frac{2E^*}{\pi\sigma_y}\sqrt{\frac{\delta_n}{R^*}}$, and $\tau_p = \frac{2E^*}{\pi\sigma_y}\sqrt{\frac{\delta_{np}}{R^*}}$.

After nonlinear regression, the model parameters are identified with $R^2 = 0.93$ as $c = 1.1$, $b = 3.0$, $m_1 = 0.2$ and $m_2 = 4.6$. As shown in Figure A1, the developed model can well reproduce the particle-scale simulation results.

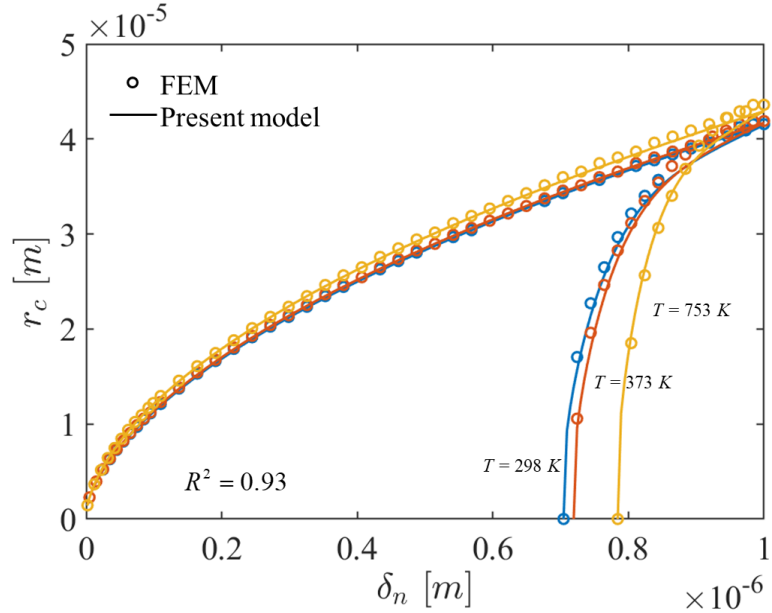

Figure A1. Validation of the present model for contact radius calculation against FEM analysis under different temperatures.

## Appendix B: Material properties

Material properties of helium and beryllium are adopted using the functions of temperature as follows [53]:

1) Gas conductivity (Helium):

$$k_g = 3.366 \times 10^{-3} T^{0.668} \text{ [W/(m K)]}$$

2) Solid Conductivity (Beryllium):

$$k_s = 430.35 - 1.1674T + 1.6044 \times 10^{-3}T^2 - 1.0097 \times 10^{-6}T^3 + 2.3642 \times 10^{-10}T^4 \text{ [W/(m K)]}$$

3) Specific heat capacity of solid:

$$C_{p,s} = 2 \times 10^{-11}(606.91 + 5.3382T - 4.1726 \times 10^{-3}T^2 + 1.2723 \times 10^{-6}T^3) \text{ [J/(kg K)]}$$

4) Linear thermal expansion coefficient:

$$\alpha = 8.43 \times 10^{-6}(T + 0.68 \times 10^{-3}T^2 - 1.18 \times 10^{-7}T^3) \, [\frac{1}{K}]$$

5) Young's modulus:

$$E = 297.0 \, e^{-0.07}(1.0 - 1.9 \times 10^{-4}(T - 293.15)) \text{ [GPa]}$$

6) Yield stress:

$$\sigma_y = 528.73 - 0.47663T + 1.0 \times 10^{-4}T^2 \text{ [MPa]}.$$